\documentclass[apj]{emulateapj}
\usepackage{graphicx}
\usepackage{natbib}

\begin{document}

\title{The ages, metallicities and star formation histories of early-type galaxies in SDSS}

\author{Raul Jimenez,\altaffilmark{1} Mariangela Bernardi,\altaffilmark{1} Zoltan Haiman,\altaffilmark{2} Ben Panter,\altaffilmark{3} Alan F. Heavens\altaffilmark{4}}

\affil{}

\altaffiltext{1}{Department of Physics and Astronomy, University of Pennsylvania, 209 South 33rd St, Philadelphia, PA 19104; raulj,bernardm@physics.upenn.edu}

\altaffiltext{2}{Department of Astronomy, Columbia University, 550 West 120th Street, New York, NY 10027, USA; zoltan@astro.columbia.edu}

\altaffiltext{3}{Max-Planck-Institut fur Astrophysik, Karl-Scwarzschild Str. 1, D-85748, Garching bei Munchen, Germany; bdp@mpa-garching.mpg.de}

\altaffiltext{4}{SUPA (Scottish Universities Physics Alliance), Institute for Astronomy, University of Edinburgh, Royal Observatory, Blackford Hill, Edinburgh EH9-3HJ, UK; afh@roe.ac.uk}

\begin{abstract} 
  We use the spectra of $\sim 22,000$ early-type galaxies, selected
  from the Sloan Digital Sky Survey, to infer the ages, metallicities
  and star formation histories of these galaxies.  We find clear
  evidence of "downsizing", i.e. galaxies with larger velocity
  dispersion have older stellar populations.  In particular, most
  early-type galaxies with velocity dispersion exceeding $200$ km
  s$^{-1}$ formed more than 90\% of their current stellar mass at
  redshift $z > 2.5$. Therefore, star formation was suppressed around
  this redshift. We also show that chemical enrichment was rapid,
  lasting 1-2 Gyr and find evidence that [Fe/H] is sub-solar.  We
  study the robustness of these results by comparing three different
  approaches: using (i) Lick absorption line indices; (ii) fitting a
  single-burst stellar population model to the whole spectrum
  (lines+continuum); and (iii) reconstructing the star formation and
  metallicity histories in multiple age-bins, providing a method to
  measure mass-weighted ages and metallicities. We find good agreement
  between the luminosity-weighted ages and metallicities computed with
  these three methods.
\end{abstract}

\keywords{galaxies: general -- galaxies: elliptical}

\section{Introduction}

It is of general interest to theories of galaxy formation to determine observationally how early-type galaxies formed as this gives clues on how feedback was regulated in these systems (e.g. \citet{Benson+02,Baugh+03,HS03,oh,Somerville+04,Nagamine+05,DeLucia+06,Bower+06,Croton+06}). The most direct route to do this is to observe the universe at different redshifts. If early-types are found up to a certain redshift, then one can determine how far back early-type galaxy formation occurred and what was the stellar mass and appearance of these galaxies. This indeed would be the best way to break the age-metallicity degeneracy, i.e. the property by which a galaxy can appear young or old due to complete lack of knowledge of its metallicity. However, as star formation becomes more intense in the past, a galaxy may hardly be classified as an early-type, thus making it difficult to find the progenitors of early-types by direct observation. Further, obtaining large samples of early-type galaxies at redshifts larger than one is very demanding observationally with the current generation of $10$m-class telescopes. Nevertheless, there has been recent progress at finding the progenitors of nearby early-type galaxies in the redshift range $1 < z < 2$ (e.g. \citet{Cimatti+03,Daddi+03,Labbe+05,Papovich+05,Caputi+06,Papovich+06,Stern+06}), however the number of progenitors found is only about one hundred. 

One alternative approach to increase significantly the statistics, and be able to do environment studies \citep{SJPH06} is to look at the local population of galaxies and try to infer from their stellar populations the ages and metallicities of the stars (e.g. \citet{Heavens+04,Panter+06}). This approach suffers from two drawbacks: first, it can only tell us how the stars were formed, and not how the dark matter and stellar material were assembled. Second, recovering information from the integrated light of the stellar populations is subject to degeneracies: for certain features of the spectral energy distribution (SED) of galaxies it is not possible to distinguish between an old stellar population with a certain metallicity and a younger one with a higher metallicity \citep{Worthey94}. On the other hand, since galaxies are nearby they are not faint and therefore large numbers can be observed with small telescopes.  Further, the use of the full spectral information significantly reduces, or removes altogether, the age-metallicity degeneracy \citep{Fanelli+87,Fanelli+92,Dunlop+96,Spinrad+97,Dorman+03,Jimenez+04}.

An interesting question is whether given enough spectral coverage and signal to noise in the SED, it is possible to recover any evolutionary information from observations of nearby early-type galaxies: the fossil record. This question has been addressed before based purely on theoretical grounds (e.g. \citet{Jimenez+04}). The question that we investigate in this paper is similar but we do it by using observations from the Sloan Digital Sky Survey (SDSS) and stellar population models that specifically include the treatment of non-solar scaled abundances. In particular, we want to know if the reconstructed star formation history derived from MOPED \citep{HJL00} is consistent with the luminosity-weighted ages and metallicities obtained with other procedures. Early-type galaxies are particularly suitable for this study since they have simpler star formation histories than disk galaxies.
 
In a previous study \citep{Heavens+04}, we recovered the star--formation histories
of a sample of SDSS galaxies, regardless of type/morphology.  In this paper, we
specifically analyze the subset of early--type galaxies.
Our main findings are as follows: the star formation of early-type galaxies can be accurately recovered with MOPED, even if the number of parameters is significant.  From the sample of early-type SDSS galaxies we find clear evidence for "downsizing", i.e. most massive galaxies form their stars first. This effect is more pronounced for early--type galaxies than for the whole galaxy population. Furthermore, we find that more than  60\% of massive galaxies have formed 90\% of their stars at $z > 2.5$. This places a strong constraint on when star formation has to be quenched in these systems. We show using hydro-dynamical models of early-type formation how this can be achieved. Furthermore, we show how chemical enrichment took place in this systems. In total we use about $40,000$ early-type galaxies, a number that is many orders of magnitude larger than any direct observations at $z \sim 2-3$.

The paper is organized as follows: \S 2 describes the sample used in this study. In \S 3 we describe the different methods used to determine the ages, star formation histories and metallicities of early-type galaxies.  \S 4 and 5 describe the comparison between the different methods. We present the star formation history in \S 6 and the metallicity history in \S 7. We draw our conclusions in \S 8. 

\section{The sample}

We have used the sample of early-type galaxies analyzed by \citet{bernardi+06}. The sample, extracted from the Sloan Digital Sky Survey \citep{York+00} in its Data Release 2 \citep{Abazajian+05}, contains over $40,000$ early-type galaxies selected for having apparent magnitude $14.5 \le r \le 17.75$ with spectroscopic parameter $eclass < 0$, which gives a PCA component corresponding to no emission lines, and $fracDev_r > 0.8$, which is a seeing-corrected indicator of morphology. In addition, only those objects with measured velocity dispersions were selected, which translates in the spectra having $S/N > 10$ in the region $4200-5800$ \AA. The sample encompasses a redshift range $0.05 < z < 0.2$, which corresponds to a maximum look-back time of 2 Gyr.

\section{The method}

\subsection{Lick Indexes}

\citet{bernardi+06} used Lick indexes \citep{Worthey94,Trager+98,Thomas+03} to derive the luminosity-weighted age and metallicity of the galaxies in the sample. In order to be able to obtain enough signal-to-noise in the spectrum to accurately measure the absorption features, they stacked the spectra that had similar properties. This reduced the sample to $925$ composite spectra. \citet{bernardi+06} use the Mgb, $<$Fe$>$, H${\beta}$ and H${\gamma_F}$ Balmer absorption lines 
and $\alpha-$enhanced stellar population models \citep{Thomas+03,Thomas+04} to derive ages and metallicities. See \citet{bernardi+06} for more details.

\subsection{Full spectral fit}

Our second technique takes advantage of the large spectral range covered by the SDSS spectra, specifically the measured flux in the region $3500 - 4000$ \AA. This region is especially useful when trying to disentangle the age-metallicity degeneracy \citep{Jimenez+04}. Furthermore, a fit to the full spectrum takes advantage of the sensitivity of the continuum to metal blanketing.  We have used the models by \citet{BC03} at a resolution of $3$\AA. We then performed a $\chi^2$ minimization  to find the best fitting model in age and metallicity, so called single stellar populations (ssp). We note that we performed the fit to individual spectra, as we find that the average $S/N$ of the spectra (above 10 per resolution element) is sufficient to yield a meaningful fit. 

\subsection{MOPED fit}

The third and final method uses the MOPED algorithm \citep{HJL00} that goes one step beyond ssp fitting and recovers in a minimally parametric way  the star formation and metallicity histories of a given galaxy (e.g. \citealt{Heavens+04}).  The recent DR3 release from the SDSS has been analyzed by \citet{Panter+06} using MOPED. It is from this study that we select those galaxies that overlap with the Bernardi catalog of early-types. This results in about $22,000$ galaxies in common. 

In brief, MOPED uses 11 bins, equally spaced  logarithmically in lookback time, to describe the star formation history of a galaxy. In each bin there are two unknowns: the strength of the burst and its metallicity, that are searched from a set of synthetic stellar population models, to produce the best fit to the observed galaxy. In addition there is a further parameter that determines the amount of dust present in the galaxy today. For this study we have used the \citet{BC03} models at $3$\AA\, resolution as the ssp describing the stellar population in each of the bins. 

One obvious concern with MOPED is that with a search in such large parameter space, degeneracies may be significant. Although this could be the case on an individual basis, \citet{Panter+03} showed that this is not the case for stellar populations with old components (age $>$ 8 Gyr). Furthermore, \citet{Panter+03,Panter+06} demonstrated that no degeneracies remain for large statistical samples. One clear advantage of MOPED is that it can provide the star formation history of the galaxy and therefore the possibility to calculate a mass-weighted age and metallicity as opposed to luminosity-weighted quantities as the other two methods. 

In the following sections we show that when one computes the luminosity-weighted age and metallicity from MOPED they agree well with those obtained from the Lick or full spectrum methods. This shows that despite the search in such a large parameter space (23 parameters), there are no significant degeneracies in the MOPED solution. 

\begin{figure}
\plotone{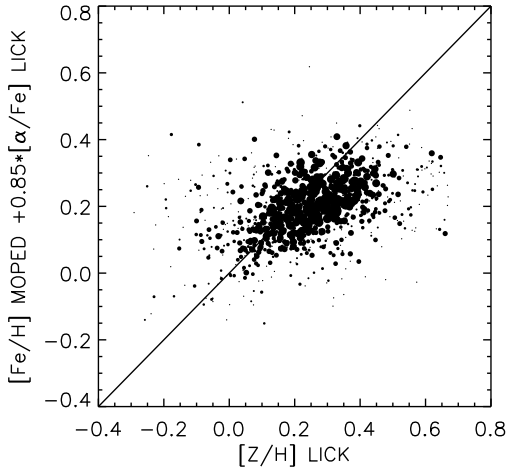}
\caption[fig:fig1]{Comparison between the luminosity-weighted metallicities obtained from the Lick indexes and the luminosity-weighted metallicity (see text) obtained from MOPED. There is good agreement between the two methods.}
\label{fig:fig1}
\end{figure}

\begin{figure}
\plotone{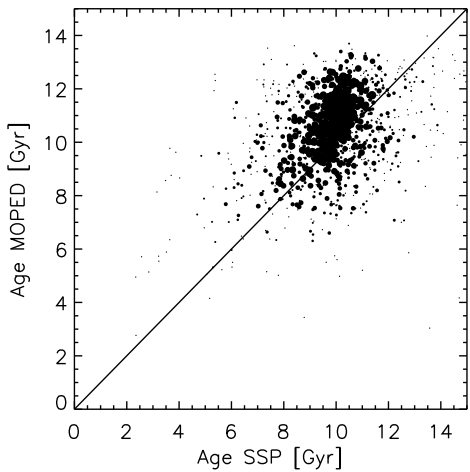}
\caption[fig:fig2]{Comparison between the luminosity-weighted age obtained with fits to the full spectrum (Age SSP) and MOPED. There is a slight trend for MOPED ages being older by about 5\% for the oldest ages. }
\label{fig:fig2}
\end{figure}

\begin{figure}
\plotone{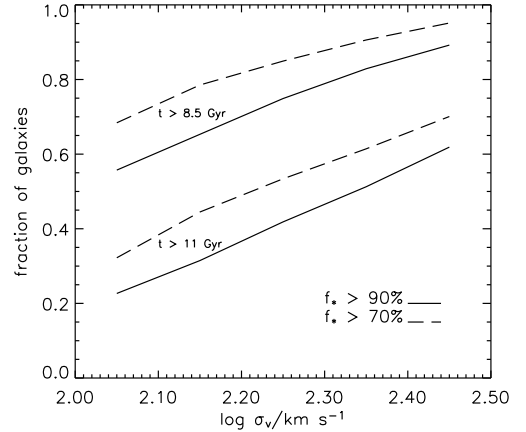}
\caption[fig:agez]{Fraction of galaxies in the SDSS sample as a function of velocity dispersion that have formed more than 90\% of their stars at a look-back time greater than 11 Gyr and 8.5 Gyr (solid lines). The dashed lines are the same but now for galaxies that have formed more than 70\% of their stars. Note the strong correlation between fraction of galaxies and velocity dispersion which is clear evidence of "downsizing" for the early-type population. Also note that for $\log (\sigma_V/{\rm km s^{-1}}) > 2.3$ more than 50\% of the galaxies have formed 90\% of their stars at $z > 2.5$. This fraction increases to 90\% of galaxies that have formed more than  90\% of their stars at $z > 1.2$. This indicates that a significant fraction of massive early-types have shut-off stars formation at $z  \sim 2.5$.}
\label{fig:agez}
\end{figure}

\section{The metallicities of early-type galaxies}

Fig.~\ref{fig:fig1} shows the comparison between the luminosity-weighted metallicity derived from the Lick indexes in \citet{bernardi+06} and from MOPED. The Lick metallicities are computed with the $\alpha-$enhanced models of  \citet{Maraston05} while the MOPED metallicities are for the solar-scaled models of \citet{BC03}. For the MOPED method we compute mass-weighted metallicities by adding up the 11 bins weighting by the luminosity of each component in the SDSS band $r$.

In order to convert the solar-scaled metallicities to the $\alpha-$enhanced scale, we use the formula by \citet{Trager+00}.  To do the comparison with the composites from \citet{bernardi+06} we have averaged the metallicities of the MOPED spectra that correspond to a given composite in the Bernardi et al. sample. 

We find good agreement between the two methods. There is a slight systematic shift at high-metallicities of about $0.1$ dex. The MOPED corrected metallicities tend to be lower than the Lick derived ones. The $1\sigma$ dispersion around the one-to-one line is of about $0.05$ dex. The metallicities for all galaxies are super-solar, with no galaxies below the solar value. The maximum metallicity is about twice the solar value. 

\section{The ages of Early-Type galaxies}

Fig.~\ref{fig:fig2} shows a comparison between the luminosity-weighted ages obtained from the single fit to the whole spectrum  and the MOPED luminosity-weighted ages (i.e. summing-up the ages of all the 11 bins weighted by their luminosity in the SDSS band $r$).

This is the more direct method to determine the extent degeneracies can play in the MOPED solution. Because the two methods use the same set of stellar populations models \citep{BC03},  the comparison is free of systematics. It gives a measurement of MOPED degeneracies because it compares a fit with just two parameters, age and metallicity, with the 23 parameters of MOPED.

The agreement is reasonable. There is a tendency for the MOPED ages greater than 10 Gyr to be older than the SSP ages by about one Gyr. However, this can be understood from the fact that the last bin of age in MOPED is quite wide (10 to 13 Gyr) and therefore the uncertainty is of the order of 20\%. As Fig.~\ref{fig:fig2} shows, and also discussed in \S~6 below, the weighted age is dominated by this last bin, so the intrinsic uncertainty in the MOPED bin dominates the error in the comparison. In the comparison we chose 11.5 Gyr as the `age' of the bin. 

The comparison with the Lick indexes ages is not as straightforward,
because the Lick index ages were calculated with $\alpha-$enhanced models.
Fully-tested $\alpha-$enhanced models that can be used to fit the full
spectrum (lines+continuum) are not yet available.

\section{The star formation histories of Early-Type galaxies}

We can now look in detail at the star formation histories provided by MOPED. In particular, we are interested to know what the typical star formation history of an early-type is and how it relates to the mass of the galaxy.

Fig.~\ref{fig:agez} shows what fraction of early-type galaxies in the SDSS have formed more than a given fraction of their current stellar mass above a given look-back time. The solid lines show the fraction of galaxies that have formed more than 90\% of their current stellar mass at  look-back times larger than 11 and 8.5 Gyr. First note the strong correlation with velocity dispersion which is a clear sign of "downsizing" for early-type galaxies, i.e. star formation occurs in more massive galaxies at earlier times (see e.g. \citealt{Cowie+99,bernardi+03,Heavens+04,kodama,thomas,tanaka,treu,van,juneau,bernardi+06,bundy,ssa,Cimatti+06,Neistein,Mouri}). Also, for galaxies with $\log (\sigma_V/{\rm km s^{-1}}) > 2.3$ more than 50\% of the SDSS early-types have formed more than 90\% of their stars at $z > 2.5$. This fraction increases to 80\% at $z > 1.2$. Clearly large galaxies ($\sigma_V > 200$ km s$^{-1}$) are shutting off their star formation at very early epochs. The dashed lines show a similar trend for a fraction in stars larger than 70\%. For the whole population of massive early-types, there is very little stellar mass being added below $z = 1.2$.

Therefore most of the star formation in massive early-types took place at $z > 2$ and in a short burst of duration $1-2$ Gyr \citep{Jimenez+99}. The hydrodynamical models in \citet{Jimenez+99} also explain the shut-off in star formation after just $1-2$ Gyr as SN feedback that heats the gas beyond the escape velocity of the dark matter halo. These models do not include any AGN feedback yet seem to reproduce well the trends for early-types in the SDSS. The clue to their success is the early assembly of the dark matter halo. Similar trends were also found in \citet{Blanton,Granato,Jimenez+05}.

Because the MOPED star formation history points to most of the star formation in early-types taking place at $t > 11$ Gyr, it is not surprising therefore that luminosity-weighted ages and mass-weighted ages are very similar \citep{HJB06}. 

\begin{figure}
\plotone{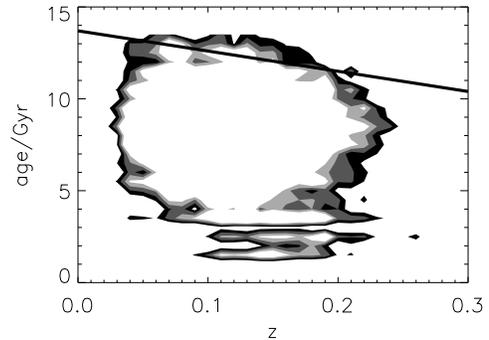}
\caption[fig:lwagez]{The age-redshift distribution for early-types in the SDSS. The white region contains 99\% of galaxies. There is a clear old edge: higher-redshift galaxies are younger than lower-redshift ones. The solid line corresponds to the age-redshift relation for the LCDM universe.}
\label{fig:lwagez}
\end{figure}

The MOPED ages are calculated over bin ages that are somewhat large (about 20\% of the age) and they are not accurate enough to see the evolution of the red envelope in detail as a function of redshift. However, it is interesting to look at the full-spectrum fit luminosity-weighted ages as a function of redshift. This assumes that the star formation took place in a single burst, but as we have seen above this is suggested by the MOPED stars formation history. Fig.~\ref{fig:lwagez} shows the age of the oldest early-type galaxies as a function of observed redshift. The plot shows a contour where the white region includes 99\% of the galaxies. For reference, the solid line is the age-redshift relation for the currently favored LCDM universe \citep{Spergel+06}. It is left for future work to provide a robust determination from these data of the Hubble constant (e.g. \citet{SVJ05}). 

\begin{figure}
\plotone{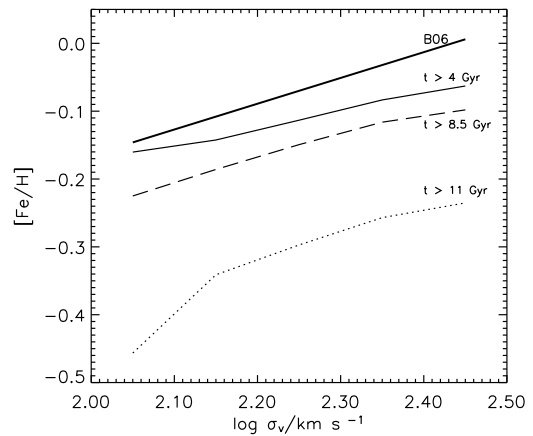}
\caption[fig:metz]{Mass-weighted Fe abundance as a function of velocity dispersion for different look-back times. Note the trend of increasing metallicity with increasing velocity dispersion at all epochs. Furthermore, there is an increase in metallicity from $t > 11$ Gyr to $t > 8.5$ Gyr and then the metallicity remains almost constant, consistent with a picture in which there is no further inflow of fresh gas. Also, we show the Fe value derived by \citet{bernardi+06} from their Lick-index study at $t=0$ Gyr.}
\label{fig:metz}
\end{figure}

\section{The metallicity histories of Early-Type galaxies} 

We now turn our attention to the evolution of metallicity in early-types. Fig.~\ref{fig:metz} shows the evolution of the mass-weighted metallicity, computed from the MOPED fit, for the SDSS early-type sample as a function of velocity dispersion and look-back time. As for the mass-weighted age, there is a correlation between metallicity and velocity dispersion. The more massive galaxies are not only the ones that contain the oldest stars but also the more metal rich. 

This is not too surprising. In fact, a simplistic model (e.g. \citet{Jimenez+99}) in which one assigns a deeper dark matter potential to the most massive galaxies will predict just that, the more massive objects are able to trap the ISM gas longer due to the higher escape velocity from the dark halo. This in turn means that gas in the more massive halos is more enriched than in the less massive ones.  

Furthermore, this model predicts that star formation will stop after $1-2$ Gyr due to the gas been heated-up above the escape velocity by SN explosions. From Fig.~\ref{fig:metz} we can see that this is the case, metal enrichment finishes at a look-back time of 8.5 Gyr ($z > 1.2$). Another interesting fact is that  the average [Fe/H] of the population is slightly below solar when mass weighted. Note that the overall metallicity is however super-solar (Fig.~\ref{fig:fig1}). This is also what one expects if star formation is quenched in $1-2$ Gyr. In this case the Type-Ia SN are not yet formed however they are the peak producers of Fe. 

\section{Conclusions}

We have presented further evidence of clear 'downsizing' for the SDSS early-type galaxies. We have also shown that these galaxies have formed a significant number of their stars at $z > 2.5$, especially the most massive ones. Furthermore, we have presented evidence that the build-up of stellar metallicity in early-type galaxies monotonically increases with time and star formation has been quenched just after $1-2$ Gyr. This can be explained by a simple model in which SN heating of the interstellar medium is sufficient to suppress further star formation by expelling it from the dark matter halo of the galaxy. \citet{Jimenez+05} noted that in the LCDM cosmology these massive halos are abundant at $z \sim 3$. They constitute the most natural candidates to host the progenitors of the SDSS early-types. The problem that remains to be solved in simulations is to suppress star formation by subsequent gas-rich mergers. This is precisely what AGN feedback provides (e.g. \citet{oh,Bower+06,Croton+06,DeLucia+06}). 

Also, we have studied in detail different methods to recover the age(s) and metallicity(s) for the integrated stellar population of the SDSS early-types. We have first compared the two most commonly-used methods to recover the luminosity-weighted age and metallicity: fitting the whole spectrum with a single-burst model, and using only Lick absorption line indices. We have shown that the two methods provide similar answers. Furthermore, we have then compared these two methods to the MOPED algorithm, which does recover a (mostly) non-parametric  star formation and metallicity history of the galaxy. Despite the larger freedom in MOPED, we have shown that the MOPED luminosity-weighted age and metallicity agree well with the above methods. Thus despite the larger number of variables MOPED uses, there does not seem to be significant degeneracies in its solution.

The large number of early-types discovered by the SDSS and the fossil record analysis provide a valuable window into the formation of this objects at $z \sim 2-3$ when the universe was only $1/5$ of its current age. We have been able to use more than $40,000$ early-type galaxies for our study. Such a large number of galaxies is not currently available at $z \sim 3$, maybe we know of a few handful of secure early-type progenitors by direct observation. As an example of the usefulness of such a large sample, in a companion paper \citep{HJB06} we use this fossil sample to constrain the evolution with redshift of the physical parameters that determine the growth of massive black holes at the centers of galaxies. With the tight constraints imposed by the fossil record, it is clear that a significant number of EROs should be detectable at such redshifts with the next generation of telescopes.
 
\section*{Acknowledgments}

The work of RJ is supported by NSF grant PIRE-0507768
and NASA grant NNG05GG01G. MB is partially supported by NASA grant 
LTSA-NNG06GC19G, and by grants 10199 and 10488 from the Space 
Telescope Science Institute, which is operated by AURA, Inc., 
under NASA contract NAS 5-26555. BP thanks the Alexander von Humboldt Foundation, the Federal Ministry of Education and Research, and the Programme for Investment in the Future (ZIP) of the German Government for funding through a Sofja Kovalevskaja award.  ZH acknowledges partial support by NASA through grants NNG04GI88G and
NNG05GF14G, by the NSF through grant AST-0307291, and
by the Hungarian Ministry of Education through a Gy\"orgy B\'ek\'esy
Fellowship. 

Funding for the Sloan Digital Sky Survey (SDSS) has been provided by the Alfred P. Sloan Foundation, the Participating Institutions, the National Aeronautics and Space Administration, the National Science Foundation, the U.S. Department of Energy, the Japanese Monbukagakusho, and the Max Planck Society. The SDSS Web site is http://www.sdss.org/.

    The SDSS is managed by the Astrophysical Research Consortium (ARC) for the Participating Institutions. The Participating Institutions are The University of Chicago, Fermilab, the Institute for Advanced Study, the Japan Participation Group, The Johns Hopkins University, the Korean Scientist Group, Los Alamos National Laboratory, the Max-Planck-Institute for Astronomy (MPIA), the Max-Planck-Institute for Astrophysics (MPA), New Mexico State University, University of Pittsburgh, University of Portsmouth, Princeton University, the United States Naval Observatory, and the University of Washington.

\end{document}